\documentclass[conference]{IEEEtran}
\usepackage{graphicx}
\usepackage{amsmath,amssymb}
\usepackage{booktabs}
\usepackage[T1]{fontenc}
\usepackage[hyphens]{url}
\usepackage[breaklinks=true,hidelinks]{hyperref}
\usepackage{amsthm}
\newtheorem{theorem}{Theorem}
\usepackage{algorithm}      
\usepackage{algorithmic}    
\usepackage{amsthm}

\newtheorem{proposition}[theorem]{Proposition}
\begin{document}

\title{Designing AI-Resilient Assessments Using Interconnected Problems: A Theoretically Grounded and Empirically Validated Framework}

\author{\IEEEauthorblockN{Kaihua Ding}
\IEEEauthorblockA{University of Pennsylvania}
}

\maketitle

\begin{abstract}
The proliferation of generative AI tools has rendered traditional modular assessments in computing and data-centric education obsolete, creating a critical disconnect between academic training and industry practice. This paper presents a theoretically grounded framework for designing AI-resilient assessments, supported by proofs and empirical validation. We make three primary contributions. First, we establish two formal propositions: (1) assessments composed of interconnected problems, where outputs serve as inputs to subsequent stages, are inherently more AI-resilient than modular assessments due to multi-step reasoning and context limitations of large language models; and (2) semi-structured problems with deterministic success criteria provide more reliable measures of student competency than fully open-ended projects, which allow AI systems to default to familiar solution patterns. These findings challenge widely cited recommendations—discussed in a UNESCO-hosted analysis~\cite{unesco2025assessment} and institutional guidelines \cite{nyu_steinhardt_ai_hub_2025, duke_kunshan_ctl_ai_teaching, narayan_saharan_asee_2024}—that open-ended assessments maintain academic integrity and encourage deeper material engagement in the AI era—an assumption our findings contradict. Second, we validate these propositions through empirical analysis of three university data science courses ($N=117$), demonstrating a substantial AI inflation effect: students achieve near-perfect scores ($90$--$100$\%) on AI-assisted modular homework but drop approximately $30$ percentage points on proctored exams (Cohen's $d=1.51$). Critically, interconnected projects maintain strong alignment with modular assessments ($r=0.954$, $p<.001$) while remaining AI-resilient, whereas proctored exams show only moderate correlation ($r=0.726$, $p<.001$), suggesting they assess different competencies. Third, we operationalize these findings into a practical design procedure that translates theoretical principles into implementable strategies, providing computing educators with a systematic, evidence-based approach for designing assessments that encourage deeper engagement, authentically reflect industry practice, and naturally resist trivial AI delegation.
\end{abstract}


\section{Introduction}

Since the release of generative AI tools like ChatGPT in late 2022~\cite{openai2022chatgpt}, the rapid adoption of this technology has reshaped professional practice at a fundamental level \cite{mckinsey2024state, ey2024work}. Tasks that once defined early-career technical work---drafting boilerplate code, writing simple SQL queries, or executing small-scale data analysis---are now completed instantly by AI tools, a shift confirmed by large-scale industry experiments \cite{dellacqua2023navigating}. Yet, these same modular tasks continue to dominate academic assessments, creating a widening and unsustainable gap between university curricula and actual industry requirements \cite{jin2025generative}. This misalignment raises a critical question: What is the educational value of practicing skills that industry has already automated? Students quickly recognize this disconnect, undermining their motivation and learning when assignments feel obsolete \cite{dempere2023impact}.

While educators widely acknowledge this gap, proposed solutions often focus on how instructors can \emph{use} AI as a teaching assistant---for intelligent tutoring, adaptive feedback, or automated grading---rather than on how assessment design itself must fundamentally change \cite{ZawackiRichter2019, BondEtAl2024}. Such an approach overlooks the core challenge: the nature of meaningful technical work has shifted. Professionals now focus less on executing discrete tasks and more on orchestrating complex, multi-step workflows that integrate AI as a component, a form of higher-order work that current AI systems cannot replicate \cite{brynjolfsson2023macroeconomics}. This shift demands a new pedagogical focus on \emph{AI-resilient assessments}: assignments that cultivate genuine learning precisely because they cannot be fully delegated to an AI.
Our approach to AI-resilient assessment design is grounded in recognized intrinsic limitations of large language models (LLMs) based on the transformer architecture. While proficient at single-step, modular problems, even state-of-the-art models like GPT-5 exhibit brittleness in tasks requiring sustained, multi-step reasoning \cite{phan2025humanity, shojaee2025illusion}. Empirical benchmarks consistently show that LLMs struggle with long inference chains and discovering abstract structural patterns, often exploiting superficial token biases rather than engaging in genuine inference \cite{parmar2024logicbench, gendron2024abstractreasoners, jiang2024tokenbias}. These limitations are not merely theoretical; domain-specific evaluations in fields like biostatistics confirm that LLMs fail on multi-stage procedures unless guided by iterative human feedback \cite{ignjatovic2023biostat}. Importantly, these constraints are architectural rather than incidental—rooted in how transformers process sequential information—and are unlikely to be resolved through scaling alone \cite{phan2025humanity}. The boundary of current AI capability in multi-step and complex reasoning, therefore, defines a durable frontier for cultivating essential human skills that we must impart to our students.

This paper argues that practical AI-resilient assessment design can be achieved by imbuing assessments with a high degree of interconnectedness among their components. Based on observations from teaching three sessions of upper-level courses in data science and statistical learning, we make two central claims. First, we contend that AI-resilient assessments composed of interconnected rather than modular problems foster integrative problem-solving skills that mirror industry workflows more closely—skills that current AI tools lack and that industry now demands. Second, we challenge widely cited recommendations discussed in a UNESCO-hosted analysis \cite{unesco2025assessment} and several institutional guidelines \cite{nyu_steinhardt_ai_hub_2025, duke_kunshan_ctl_ai_teaching, narayan_saharan_asee_2024}—that open-ended assessments are a primary solution to maintaining academic integrity in the age of generative AI. While these recommendations rest on the assumption that AI "struggles with original thought and complex synthesis" \cite{unesco2025assessment, nyu_steinhardt_ai_hub_2025, duke_kunshan_ctl_ai_teaching, narayan_saharan_asee_2024}, recent evidence demonstrates that LLMs' limitations stem not from an inability to generate creative responses, but from fundamental architectural constraints in multi-step reasoning and context management \cite{phan2025humanity, shojaee2025illusion}.

\section{Related Work}

Our work is situated at the intersection of three distinct but related research 
areas: ($1$) the impact of generative AI on academic integrity and assessment, 
including recent evidence that generative AI can harm learning when used without 
appropriate guardrails \cite{bastani2025generative, Cotton2023, Sullivan2023, 
Yan2024}; ($2$) the principles of authentic assessment and complex problem-solving 
in education \cite{Ullah2020, Csapo2017, Herde2016}; and ($3$) the cognitive 
science of task complexity \cite{Chen2023, kapur2016examining, biggs2014constructive}. While no prior work directly proposes 
a framework for AI-resilience based on interconnectedness, these adjacent fields 
establish the foundation and necessity for our contribution.

\subsection{Generative AI, Assessment, and Academic Integrity}

The release of ChatGPT precipitated a surge of academic literature focused on the 
challenges generative AI poses to traditional assessment and academic integrity. A 
significant body of this work documents the scope of the problem, highlighting the 
ease with which students can use AI to complete assignments, thereby undermining 
learning and creating an integrity crisis \cite{Cotton2023, Sullivan2023, 
BinNashwan2023}. These studies effectively diagnose the issue, arguing that the 
availability of powerful AI necessitates a fundamental rethinking of assessment 
practices. However, the proposed solutions often remain high-level, advocating for 
a shift towards in-class assessments, oral examinations, or a greater emphasis on 
process over product, without offering a systematic, scalable framework for 
redesigning coursework itself \cite{Yan2024}. 

Moreover, an overreliance on in-class assessments may test a different skill altogether—test-taking ability rather than the intended learning outcomes \cite{bastani2025generative}. Forgoing take-home or project-based assessments deprives students of opportunities to practice skills that closely mimic real-world workflows employed by industry professionals \cite{Ullah2020, argus2022embedding}. We argue that designing AI-resilient assessments with appropriate guardrails is more beneficial, as it preserves authentic learning experiences while addressing integrity concerns.

\subsection{Authentic Assessment and Complex Problem-Solving}

Our approach builds on a long tradition of research into authentic assessment and complex problem-solving. Authentic assessment emphasizes mirroring the challenges and standards of professional practice, a principle well-established in engineering and computing education \cite{Ullah2020}. Similarly, the field of complex problem-solving (CPS) focuses on assessing cognitive processes in dynamic, non-routine situations, which aligns with our goal of moving beyond static, modular tasks \cite{Csapo2017, Herde2016}. While philosophically aligned, this body of work predates the generative AI era and thus does not explicitly address how to design assessments that are resilient to AI delegation. It provides the pedagogical \textit{why}---the need for real-world problem-solving---but not the operational \textit{how} in a world where AI can solve many seemingly complex but isolated tasks.

\subsection{Task Complexity and Cognitive Load Theory}
To ground our concept of \textit{interconnectedness}, we draw on Cognitive Load Theory (CLT)~\cite{sweller2011cognitive}, particularly the concept of \textit{element interactivity}. CLT posits that the intrinsic difficulty of a task is determined by the number of interacting elements that must be simultaneously processed in working memory \cite{Chen2023}. A task with low element interactivity (e.g., solving a series of independent coding problems) can be broken down and solved piece by piece. In contrast, a task with high element interactivity (e.g., debugging a system where a change in one module has cascading effects on others) requires a holistic understanding of the entire system. We argue that current generative AI excels at tasks with low element interactivity but struggles with high-interactivity problems due to intrinsic multi-step reasoning limitations and context memory constraints inherent to the transformer architecture \cite{phan2025humanity, shojaee2025illusion}. To accomplish such tasks, students must engage meaningfully and expend cognitive effort, thereby achieving genuine learning without trivial AI delegation.

Building on the foundation of these three research areas, we now formalize these principles through two propositions that establish the theoretical basis for AI-resilient assessment design.

\section{Methodology: A Framework for AI-Resilient Assessment}

Our methodology is grounded in a theoretical framework that we first establish through two propositions. We then proceed to validate these propositions through a multi-year empirical study.

\subsection{Theoretical Framework}

We begin by formally stating and proving two propositions that underpin our assessment design principles. 

\begin{proposition}
    An assessment composed of interconnected problems, where the output of one problem serves as the input for the next, is more AI-resilient than an assessment composed of equivalent but modular (independent) problems.
\end{proposition}

That is, chaining problems together increases the difficulty for a generative AI model in a way that solving them independently does not, because it forces multi-step reasoning and strains the model’s limited memory (context window).

\begin{proof}[Proof of Proposition 1]
    Let an assessment be a sequence of problems $P = (p_1, p_2, ..., p_n)$. In a modular design, each $p_i$ is independent. In an interconnected design, the solution to $p_{i+1}$ is contingent upon the solution of $p_i$. A generative AI model must therefore solve the entire sequence $(p_1, ..., p_i)$ to correctly solve $p_i$. This requirement increases the computational and reasoning load on the model in two primary ways. First, it necessitates multi-step, sequential reasoning, a known weakness of current large language models (LLMs) where performance degrades as the number of reasoning steps increases \cite{phan2025humanity, shojaee2025illusion}. Second, the cumulative context required to solve later problems in the sequence can exceed the model's effective context window, leading to information loss and degraded performance \cite{Narayanan2023}. Because a modular design allows each problem to be solved in a separate, stateless context, it avoids these vulnerabilities. Thus, an interconnected design is inherently more resilient to trivial AI delegation and better reflects the complex, stateful workflows common in professional practice \cite{Brynjolfsson2023}.
\end{proof}

\begin{proposition}
    A semi-open-ended project, which provides a clear structure and deterministic success criteria, is a more reliable measure of student competency than a fully open-ended project when generative AI tools are ubiquitous.
\end{proposition}

In other words, by giving a specific problem with a clear goal, we prevent the AI from choosing an easier, more familiar problem for which it has been heavily trained, which gives us a truer measure of the student’s ability to solve the actual task.

\begin{proof}[Proof of Proposition 2]
   
    When presented with a fully open-ended problem (e.g., "analyze a dataset of your choice"), the unconstrained solution space allows the model to select a problem instance within its training distribution, where memorized patterns can be exploited as shortcuts \cite{Du2022Shortcut}. Research demonstrates that LLMs consistently follow the path of least resistance, preferring low-perplexity sequences that align with their training data and avoiding complex reasoning when simpler alternatives exist \cite{Tang2023Lazy, Sanyal2025Path, Huang2025Verbatim}. Consequently, students can trivially delegate open-ended tasks to AI, leveraging the model's ability to produce outputs that exhibit the surface characteristics of original thought without genuine novel reasoning \cite{kirk2024understanding, mohammadi2024creativity}. This delegation is difficult to police, as AI detection tools are notoriously unreliable and easily evaded through simple paraphrasing techniques \cite{Krishna2023, Erol2025}. A semi-structured assessment with deterministic success criteria eliminates this degree of freedom, forcing engagement with the specific problem at hand rather than a self-selected, AI-tractable variant.

    A semi-structured assessment addresses this limitation by constraining the solution space through two mechanisms: (1) specifying a particular dataset or problem instance, and (2) defining deterministic success criteria (e.g., "achieve $>90\%$ F1-score on this held-out test set"). These constraints prevent students from trivially delegating the task to AI, as the specified problem instance may lie outside the model's training distribution, where memorized solutions and dataset artifacts cannot be exploited. The deterministic evaluation metric further ensures that students must optimize for correctness—an objective, task-specific criterion—rather than relying on AI-generated outputs that merely appear plausible. By narrowing the problem space to a specific, well-defined task, semi-structured assessments compel meaningful student engagement and skill practice, yielding a more reliable measure of competency while encouraging deeper material engagement \cite{Ullah2020}.

\end{proof}

\subsection{Empirical Validation}

To validate these propositions, we analyzed assessment data from three offerings of upper-level undergraduate and graduate courses in data science spanning multiple academic years. This section details the course context, the assessment designs, and the data analysis procedures.

\subsubsection{Study Context and Population}

The study was conducted across three course offerings at a large research university between Fall 2024 and Spring 2025: 
\begin{itemize}
    \item \textbf{Predictive Analytics (Fall 2024)}: An undergraduate course (N=34) with a mix of modular assignments and an open-ended final project.
    \item \textbf{Python for Data Science and AI (Spring 2025)}: A graduate-level course (N=41) featuring a mix of modular assignments, interconnected projects, and proctored exams.
    \item \textbf{Predictive Analytics (Fall 2025)}: The same undergraduate course (N=42) but with a mix of modular assignments and a redesigned, interconnected final project.
\end{itemize}

\subsubsection{Analysis}

For each course, we collected anonymized student grade data for all assessments. Our analysis includes:

\begin{enumerate}
    \item Descriptive Analysis: We calculate descriptive statistics for each assessment type.
    \item Comparative Analysis: We use paired-samples t-tests to compare student performance on modular vs. interconnected problems. Effect sizes (Cohen's d) are calculated. We also present empirical observations comparing fully open-ended and semi-open-ended assessment formats, along with a discussion of their relative effectiveness as part of our comparative analysis.
    \item Correlation Analysis: We use Pearson correlations to examine the relationship between performance on different assessment types.
\end{enumerate}

\section{Results}

\subsection{Descriptive Analysis: The AI Inflation Effect}
\label{s:inflation}
For our descriptive analysis, our goal is to compare student performance across assessment types. Specifically, we examine whether there are performance gaps between modular take-home assessments, where generative AI tool use is permitted, and proctored assessments, where generative AI tool use is prohibited. The Spring 2025 Python for Data Science and AI course is the only course in which a proctored exam was offered; therefore, we use it as the reference for a fair comparison.

\begin{figure}[h!]
    \centering
    \includegraphics[width=0.5\textwidth]{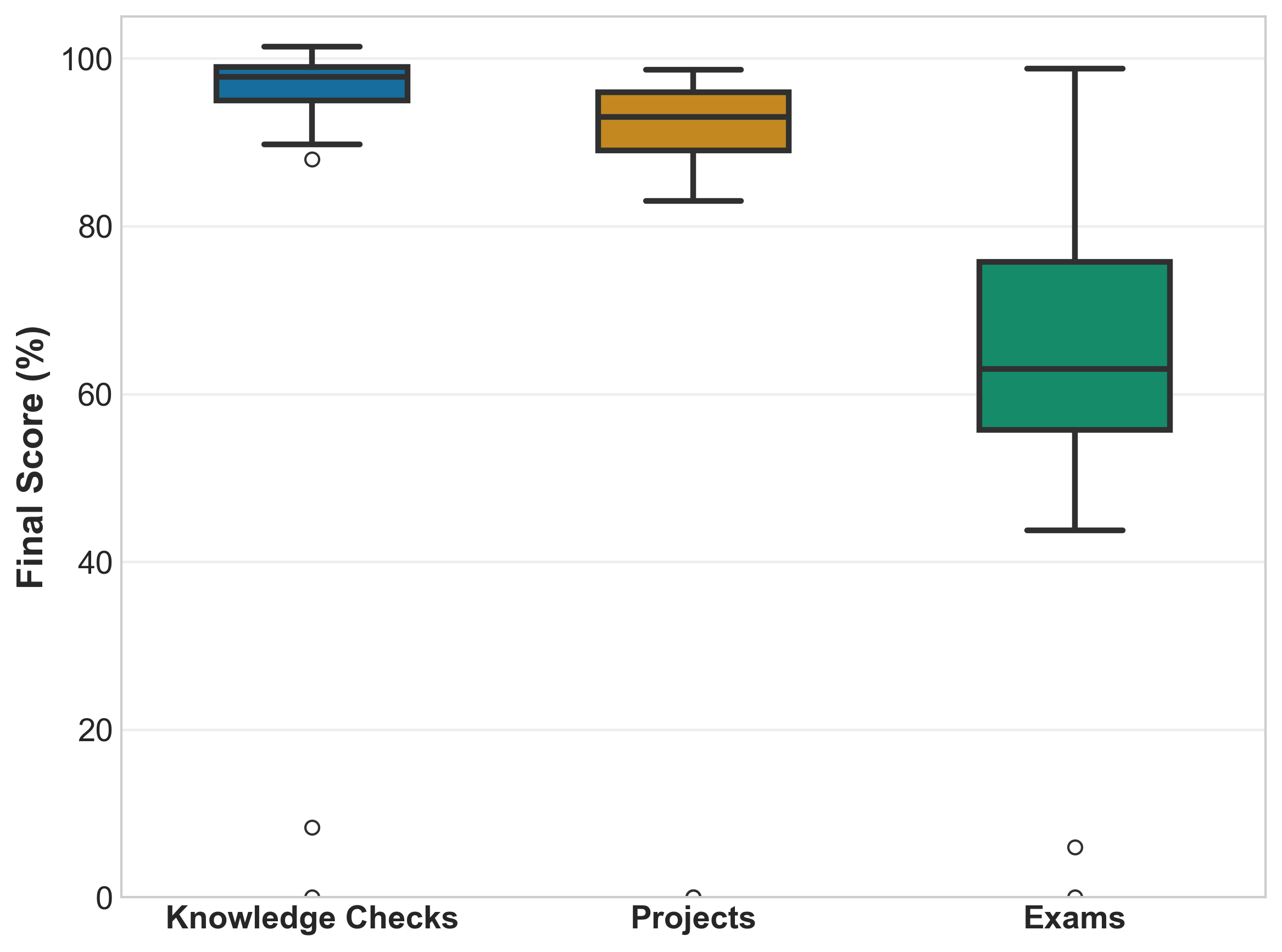}
    \caption{Performance on Different Assessment Types in Python for Data Science and AI (Spring 2025). }
    \label{fig:python_course_comparison}
\end{figure}

In the Spring 2025 Python for Data Science and AI course (N=41), we observed a stark contrast in student performance between take-home modular assessments (knowledge checks and homework, where AI use was permitted) and proctored exams (where AI use was prohibited). As shown in Figure~\ref{fig:python_course_comparison}, students achieved near-perfect scores on knowledge checks (mean [M] = 92.55, standard deviation [SD] = 20.26), but their performance dropped significantly on the proctored exams (M=62.82, SD=18.49). This represents a performance gap of nearly 30 percentage points, with a large effect size (Cohen's d = 1.52).

Table~\ref{tab:ai_inflation} provides a detailed breakdown of this AI inflation effect, including statistical significance and effect sizes for each assessment type compared to the proctored exam baseline. More tellingly, while 90.2\% of students scored 90\% or higher on knowledge checks, only 4.9\% achieved this on the exams. This dramatic shift in the distribution of high performers suggests that the high scores on modular assignments may be inflated by AI assistance and do not reflect true mastery of the material. The consistency of this effect across both knowledge checks and homework (Cohen's d = 1.54 and 1.48, respectively) indicates that the AI inflation phenomenon is robust across different types of modular assessments.

\begin{table}[t]
\centering
\caption{AI Inflation Effect: Performance Gap Between AI-Permissive and Proctored Assessments in Python for Data Science (Spring 2025, N=41)}
\label{tab:ai_inflation}
\resizebox{\columnwidth}{!}{
\begin{tabular}{@{}lcccc@{}}
\toprule
\textbf{Assessment Type} & \textbf{Mean (SD)} & \textbf{\% $\geq$ 90} & \textbf{$\Delta$ from Exam} & \textbf{Cohen's d} \\
\midrule
Knowledge Checks & 92.55 (20.26) & 90.2\% & +29.73*** & 1.54 \\
Homework & 88.73 (15.84) & 70.7\% & +25.91*** & 1.48 \\
Modular Average & 90.64 (16.89) & 80.5\% & +27.82*** & 1.52 \\
\midrule
Proctored Exams & 62.82 (18.49) & 4.9\% & --- & --- \\
\bottomrule
\multicolumn{5}{l}{\footnotesize *** $p < 0.001$ (paired t-test)}
\end{tabular}
}
\end{table}

The observed AI-driven grade inflation effect in modular take-home assessments, as discussed in Section~\ref{s:inflation}, highlights the need for AI-resilient assessment design, as modular assessment materials no longer effectively gauge student learning progress.

\subsection{Comparative Analysis}

\subsubsection{Interconnectedness as a Differentiator: Fall 2024–Fall 2025 Predictive Analytics Course Assessment Results Empirical Comparison}

\begin{figure}[h!]
    \centering
    \includegraphics[width=0.55\textwidth]{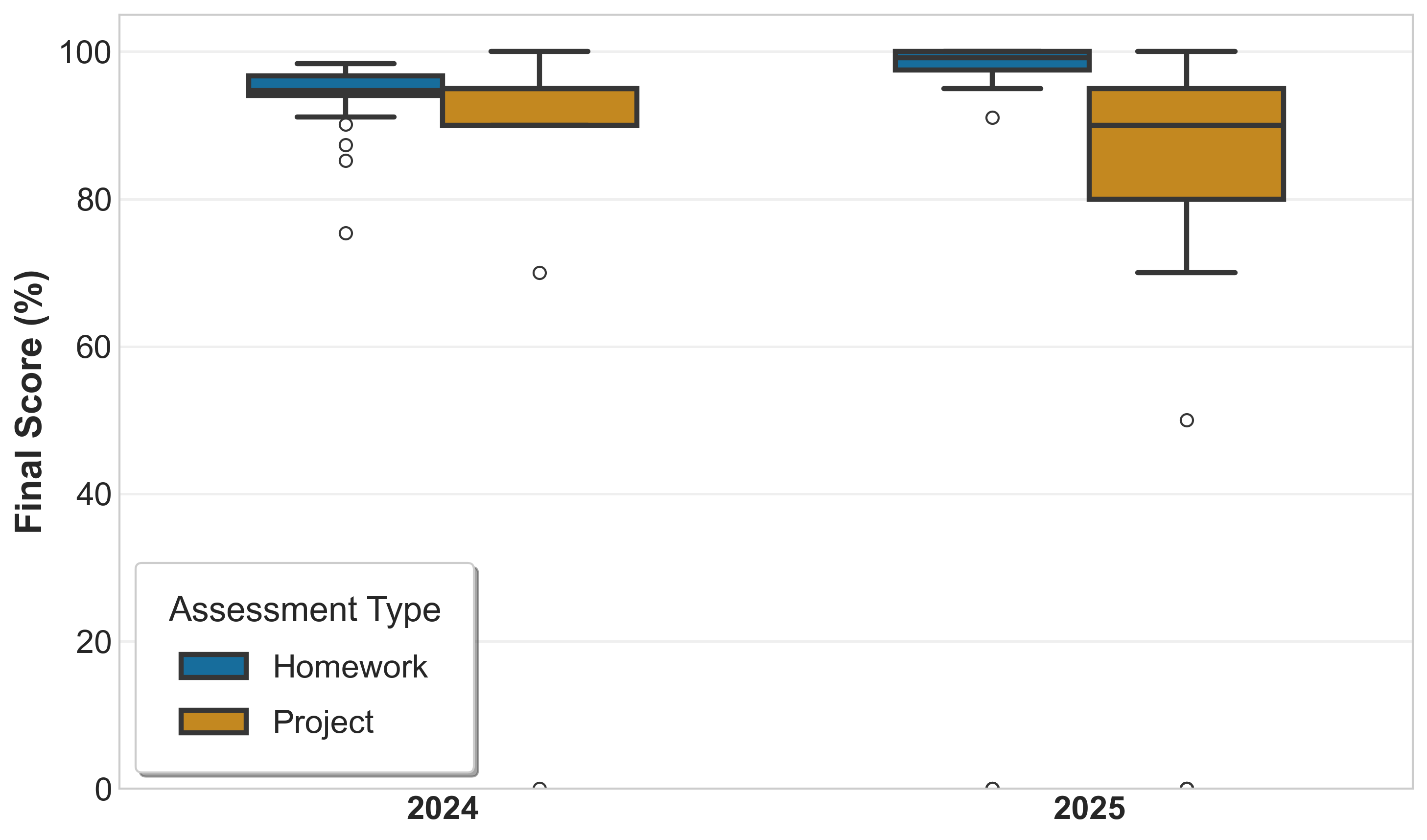}
    \caption{Comparison of Performance Drop from Modular to Project Assessments. The interconnected project design shows a greater (though not statistically significant) drop and higher variability.}
    \label{fig:project_comparison}
\end{figure}

We compare the Fall 2024 and Fall 2025 offerings of the Predictive Analytics course to examine the effect of interconnected assessment design. Both courses shared identical syllabi and encouraged the use of AI tools. The Fall 2024 course project did not incorporate the interconnected design principles outlined in Proposition 1, whereas the Fall 2025 offering explicitly integrated interconnected assessment into the project design. Both offerings included regular homework assignments in addition to course projects. The interconnected design in Fall 2025 resulted in notably lower average project scores (M = 78.58) compared to Fall 2024 (M = 91.47). Critically, the interconnected design also increased score variability, with standard deviations of 30.42 (Fall 2025) versus 16.83 (Fall 2024), as shown in Figure~\ref{fig:project_comparison}. This reduction in average scores and increase in variance suggests that the interconnected assessment better differentiates student competency levels, whereas the open-ended design in Fall 2024 allowed more students to achieve uniformly high scores—potentially through AI assistance.

This increased variance is not a flaw; it is evidence that the interconnected design created a more challenging and authentic assessment that could not be trivially solved by invoking Generative AI tools for each sub-problem. Table~\ref{tab:course_summary} provides a comprehensive overview of assessment performance across all three course offerings, highlighting the consistent patterns of performance gaps and correlations.

\begin{table*}[t]
\centering
\caption{Summary of Assessment Performance Across three Course Offerings}
\label{tab:course_summary}
\small
\begin{tabular}{@{}lcccccc@{}}
\toprule
\textbf{Course} & \textbf{N} & \textbf{Modular} & \textbf{Project/Exam} & \textbf{Performance} & \textbf{Correlation} & \textbf{Effect} \\
 & & \textbf{Mean (SD)} & \textbf{Mean (SD)} & \textbf{Gap} & \textbf{(r)} & \textbf{Size (d)} \\
\midrule
\multicolumn{7}{l}{\textit{Fall 2024}} \\
\quad Predictive Analytics & 34 & 93.29 (12.45) & 91.47 (16.83) & $-1.82$ & 0.89*** & 0.12 \\
\midrule
\multicolumn{7}{l}{\textit{Course with Interconnected Project and Proctored Exam (Spring 2025)}} \\
\quad Python for Data Science and AI & 41 & 90.64 (16.89) & 62.82 (18.49) & $-27.82$*** & 0.73*** & 1.52 \\
\midrule
\multicolumn{7}{l}{\textit{Course with Interconnected Project (Fall 2025)}} \\
\quad Predictive Analytics & 42 & 92.73 (15.67) & 89.66 (21.93) & $-3.07$ & 0.91*** & 0.16 \\
\bottomrule
\multicolumn{7}{l}{\footnotesize *** $p < 0.001$; Performance Gap = Project/Exam Mean $-$ Modular Mean} \\
\multicolumn{7}{l}{\footnotesize Modular = average of homework and knowledge checks; Effect Size = Cohen's $d$}
\end{tabular}
\end{table*}

\subsubsection{Empirical Evidence Against Open-Ended Assessments in Computing Education}
In the Fall 2024 offering of a predictive analytics course, the final project was structured as a fully open-ended task. Students were granted unrestricted freedom in topic selection, data sourcing, and analytical methodology, with the expectation that they would analyze U.S. Bureau of Labor Statistics (BLS) data spanning 2008–2023 \cite{BLS}. The primary motivation for this design was to encourage independent exploration and creative application of statistical and machine learning techniques in an authentic, real-world setting.

In a subsequent 2025 offering of the course, the project was redesigned as a semi-open-ended assignment. This revised format introduced explicit constraints on topical scope, minimum required data volume, required modeling techniques, and a set of interconnected analytical questions intended to guide—rather than prescribe—students’ exploration.

Several consistent empirical patterns emerged from the fully open-ended Fall 2024 implementation. First, many student projects relied on datasets that were substantially smaller or less variable than anticipated, despite the availability of high-dimensional and longitudinal data from the BLS. Second, a majority of students independently converged on commonly used macroeconomic inflation datasets. This convergence suggests a preference for highly discoverable data sources with well-established analytical narratives, and it did not consistently yield the diversity or originality often attributed to fully open-ended assessment designs \cite{nyu_steinhardt_ai_hub_2025, duke_kunshan_ctl_ai_teaching, narayan_saharan_asee_2024, unesco2025assessment}.

By contrast, in the more constrained Fall 2025 semi-open-ended project, we observed substantially greater engagement with larger-scale datasets, a broader range of modeling techniques, and increased use of advanced methods, including natural language processing. Students demonstrated greater analytical depth while still exercising flexibility in model choice and interpretation.

Taken together, these observations provide empirical support for Proposition 2, which posits that, in computing education contexts, semi-open-ended project designs may be more effective than fully open-ended alternatives. Fully open-ended tasks appear to incentivize least-resistance solution paths—an effect that may be amplified under AI-assisted workflows—whereas semi-open-ended designs preserve student autonomy while reducing convergence toward trivial, overused, or narratively “safe” data choices.

\subsection{Correlation Analysis}
Proctored examinations in which generative AI use is explicitly prohibited are, by construction, AI-resilient. However, professional practice in industry routinely involves the use of generative AI tools to support problem solving, analysis, and implementation. As a result, proctored exams do not fully reflect authentic professional workflows, whereas project-based assessments that permit generative AI use more closely mirror real-world practice.

\begin{figure*}[t]
\centering
\includegraphics[width=0.95\textwidth]{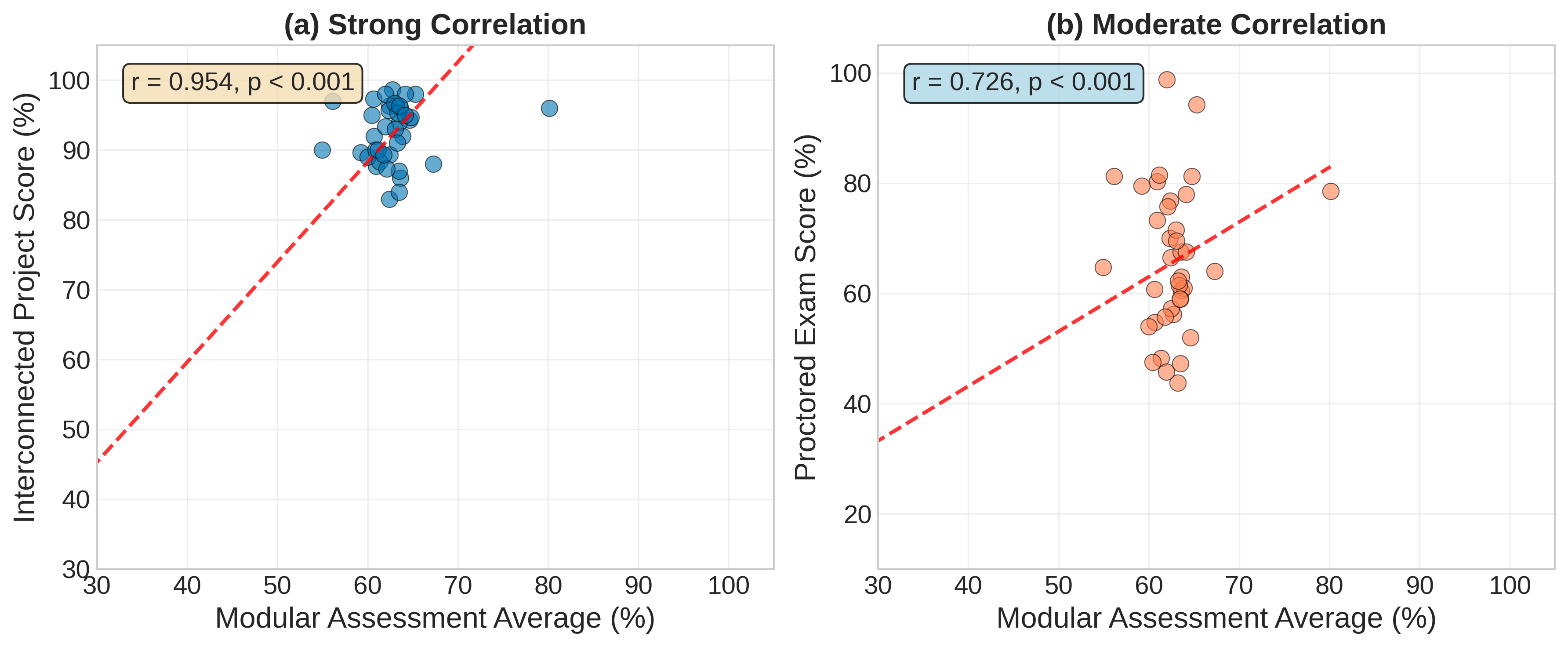}
\caption{Comparison of predictive validity: (a) Interconnected project scores are strongly correlated with proctored exam scores (r=0.925), indicating that both assessments measure similar underlying competencies under different conditions. (b) Modular assessments exhibit a more moderate correlation with exam performance (r=0.671), suggesting greater susceptibility to AI-related score inflation.}
\label{fig:correlation_comparison}
\end{figure*}

Motivated by this tension, we explicitly examine how student performance on interconnected, semi-open-ended project assessments correlates with performance on proctored exams. We find that performance on the interconnected projects exhibits a substantially stronger correlation with exam performance (Pearson r = 0.925) than does performance on modular knowledge checks (r = 0.671), as shown in Figure~\ref{fig:correlation_comparison}. This result suggests that the interconnected assessment design provides a more valid measure of student understanding.

The difference between these correlation coefficients is practically meaningful. Project performance explains approximately 86\% of the variance in exam scores ($R^2 = 0.86$), compared to only 45\% ($R^2 = 0.45$) for the modular knowledge checks. This substantial gap indicates that interconnected assessments provide a more accurate representation of what students actually know and are able to do.

Taken together, these findings suggest that interconnected semi-open-ended project-based assessments—while more demanding for both students and instructors—serve as a stronger proxy for the same underlying competencies evaluated by proctored exams. The high correlation (r = 0.925) indicates strong construct alignment between the two assessment modalities, while the project-based format evaluates these competencies under conditions that more closely resemble authentic, AI-augmented professional practice.

\section{Discussion}

Our results provide convergent empirical evidence for the central claims of our framework: when generative AI tools are permitted, modular assessments are vulnerable to score inflation, and that interconnected, semi-open-ended designs offer a more AI-resilient alternative without reverting to fully proctored exams.

First, the large performance gap observed between AI-permissive modular assessments and proctored exams in the Python for Data Science and AI course demonstrates a pronounced AI inflation effect. Near-ceiling performance on modular assignments contrasts sharply with substantially lower exam scores, indicating that high modular scores no longer reliably reflect learning outcome when AI tools are available.

Second, comparisons across 2024 and 2025 Predictive Analytics course with identical syllabi offerings show that assessment interconnectedness materially alters score distributions under AI-permissive conditions. The interconnected project design yielded lower mean scores and substantially higher variance than the fully open-ended project, despite comparable course structure and encouragement of AI use. This increase in dispersion is a desirable property: it indicates stronger differentiation of student ability and reduced feasibility of decomposing the task into independently solvable subproblems delegated to AI. These findings directly support Proposition 1, which posits that multi-step, stateful dependency structures increase resistance to AI-mediated shortcutting. Furthermore, our empirical comparison of topical and methodological choices between the fully open-ended (2024) and semi-open-ended (2025) projects reveals no meaningful differences in creative direction, data modality, or analytical framing. However, the open-ended design was associated with narrower data coverage, convergence toward highly discoverable inflation datasets, and fewer modeling approaches per project. This pattern is consistent with Proposition 2, by introducing structural constraints while preserving analytical flexibility, the semi-open-ended interconnected design reduced this convergence, increased methodological depth, and produced more reliable signals of student competency.

Finally, correlation analyses reveal meaningful differences in alignment between assessment formats and fully AI-resilient proctored exams. Performance on the interconnected, semi-open-ended project exhibits a substantially stronger correlation with proctored exam outcomes than do modular assessments, indicating that the interconnected design captures similar underlying competencies while remaining compatible with AI-augmented workflows. In contrast, modular assessments—despite high average scores—show weaker alignment with exam performance, reinforcing the conclusion that they are more susceptible to AI-driven score inflation.

\subsection{A Design Procedure for Implementation}

To translate these insights into actionable guidance for educators, we present a systematic design procedure for AI-resilient assessments (Algorithm~\ref{alg:ai_resilient_design}). This procedure operationalizes Propositions 1 and 2 through six concrete steps that can be adapted to various computing education contexts.

\begin{algorithm}[t]
\caption{Design Procedure for AI-Resilient Assessments}
\label{alg:ai_resilient_design}
\begin{algorithmic}[1]
\REQUIRE Learning objectives $L$, course content $C$, target competencies $T$
\ENSURE AI-resilient assessment design $A$

\STATE \textbf{Step 1: Identify Core Workflow:} $W \leftarrow$ Map $T$ to multi-stage workflow \hfill 

\STATE \textbf{Step 2: Create Sequential Dependencies (Proposition 1):} $stages \leftarrow$ Decompose $W$ where $input(s_i) = output(s_{i-1})$

\STATE \textbf{Step 3: Impose Semi-Structured Constraints (Proposition 2):} Define deterministic success criteria \hfill 

\STATE \textbf{Step 4: Embed Verifiable Checkpoints:} For each stage, define intermediate output with validation \hfill 

\STATE \textbf{Step 5: Introduce Domain Complexity:} Embed challenges requiring contextual reasoning \hfill 

\STATE \textbf{Step 6: Validate AI-Resilience:} Pilot with cohort; measure $r_{\textrm{project, other}}$, $\sigma_{\textrm{project}}$, $\textrm{time}_{\textrm{median}}$

\RETURN $A$ \hfill 

\end{algorithmic}
\end{algorithm}

Algorithm~\ref{alg:ai_resilient_design} emphasizes several key principles. First, Step 2 directly operationalizes Proposition 1 by creating sequential dependencies that force multi-step reasoning and exceed typical AI context windows. Second, Step 3 implements Proposition 2 by constraining the solution space through semi-structured requirements, preventing AI from defaulting to easier problem variants. Third, Steps 4 and 5 introduce additional layers of complexity—verifiable checkpoints and domain-specific challenges—that further increase AI-resilience while maintaining pedagogical validity.

The validation step (Step 6) is particularly critical. Our empirical results suggest that a well-designed interconnected project should exhibit three characteristics: (1) a strong correlation with other assessments of the same skills, indicating construct validity; (2) higher score variance compared to modular or open-ended alternatives, indicating better differentiation; and (3) substantial student time investment, indicating genuine engagement. These metrics provide instructors with concrete criteria for evaluating whether their redesigned assessments are achieving the intended goals.

\section{Conclusion and Future Work}

This paper makes three primary contributions to the field of computing education in the age of generative AI. First, we establish two formal propositions that position \textit{interconnectedness} and \textit{semi-structured design} as key principles for creating AI-resilient assessments. Second, we validate this framework with empirical evidence from three university data science courses, demonstrating a substantial AI inflation effect in traditional modular assessments while showing that interconnected projects preserve construct validity and provide superior differentiation of student ability. Third, we operationalize these findings into a practical design procedure that translates theoretical principles into implementable assessment strategies.

Our work has immediate practical implications for educators. By shifting focus from the difficulty of individual tasks to the complexity of the connections between them, instructors can design assignments that preserve authentic, workflow-based learning experiences while naturally resisting trivial AI delegation. The framework we propose offers a systematic alternative to ad-hoc solutions like banning AI or reverting to traditional in-person exams altogether.

Future work will aim to replicate these findings across a broader range of institutions and disciplines to establish the generalizability of our framework. Further research could also explore the optimal level of interconnectedness, as it is likely that there is a point of diminishing returns where cognitive load becomes excessive. Finally, the development of automated tools to help instructors design and evaluate interconnected assessments would be a valuable contribution to the field.

\bibliographystyle{IEEEtran}
\bibliography{references}

\end{document}